%% file: main_ieee.tex
\documentclass[conference]{IEEEtran}
\IEEEoverridecommandlockouts
\usepackage{cite}
\usepackage{amsmath,amssymb,amsfonts}
\usepackage{algorithmic}
\usepackage{graphicx}
\usepackage{textcomp}
\usepackage[dvipsnames]{xcolor}
\def\BibTeX{{\rm B\kern-.05em{\sc i\kern-.025em b}\kern-.08em
    T\kern-.1667em\lower.7ex\hbox{E}\kern-.125emX}}

\usepackage{listings}
\usepackage{physics}
\usepackage{AEmilia}
\usepackage{algorithm}
\usepackage{adjustbox}
\usepackage{subcaption}
\usepackage{epstopdf}
\usepackage{easyReview}
\usepackage{dirtytalk}
\usepackage{balance}

\usepackage[british]{babel}

\usepackage{svg}
\usepackage{tcolorbox}

\usepackage[spaces,hyphens]{xurl}

\usepackage{latexsym,tabularx,multirow,multicol,xspace,tikz}
\usepackage{makecell}
\usepackage{booktabs}
\usepackage{ltxtable}
\usepackage{colortbl}
\usepackage{arydshln}
\usepackage{rotating} 

\newcommand{\easier}{\emph{EA\-SIER}\xspace}
\newcommand{\nsga}{\ensuremath{\mathtt{NSGA-II}}\xspace}

\newcommand{\pas}{\textsc{\#pas}\xspace}
\newcommand{\perfq}{\textsc{perfQ}\xspace}
\newcommand{\achanges}{\textsc{archDist}\xspace}
\newcommand{\reliability}{\textsc{reliability}\xspace}

\newcommand{\ie}{\emph{i.e.,}\xspace}
\newcommand{\eg}{\emph{e.g.,}\xspace}

\newcommand{\etal}{\emph{et~al.}\xspace}
\newcommand{\secref}[1]{Section~\ref{#1}\xspace}
\newcommand{\figref}[1]{Figure~\ref{#1}\xspace}

\newcommand{\tabref}[1]{Table~\ref{#1}\xspace}

\newcommand{\ttcomp}{\textbf{11} UML Components\xspace}
\newcommand{\ttnode}{\textbf{11} UML Nodes\xspace}
\newcommand{\ttusecase}{\textbf{3} UML Use Cases\xspace}

\usepackage{todonotes}
\xdefinecolor{gray95}{gray}{0.80}
\xdefinecolor{gray25}{gray}{0.95}

\usepackage{framed}
\colorlet{shadecolor}{blue!10}

\newboolean{showcomments}

\setboolean{showcomments}{true}

\ifthenelse{\boolean{showcomments}}
  {\newcommand{\nb}[2]{
    \fbox{\bfseries\sffamily\scriptsize#1}
    {\sf\small$\blacktriangleright$\textit{#2}$\blacktriangleleft$}
   }
  }
  {\newcommand{\nb}[2]{}
  \setreviewsoff
  }
\newcommand\DDP[1]{\textcolor{Plum}{\nb{DANIELE}{#1}}}

\newcommand\MT[1]{\textcolor{MidnightBlue}{\nb{MICHELE}{#1}}}
\newcommand\VIC[1]{\textcolor{Red}{\nb{VITTORIO}{#1}}}

\usepackage{mdframed}
\usepackage{ntheorem}
\theoremstyle{nonumberplain}
\newmdtheoremenv[%
  linecolor=blue,
  linewidth=0pt,
  rightline=false,
  leftline=false]{figrev}{}

\newcommand\RQ[2]{
\noindent\fbox{%
    \parbox{.99\linewidth}{%
        \textbf{#1:} {#2}
    }%
}
\vspace{0.01cm}
}

\newcommand\answerRQ[1]{
        #1
\vspace{0.01cm}
}

\begin{document}

\title{On the impact of Performance Antipatterns in multi-objective software model refactoring optimization
\thanks{This research was supported by the AIDOaRt project (ECSEL-JU program - grant agreement n.101007350).}
}


\author{\IEEEauthorblockN{Vittorio Cortellessa\IEEEauthorrefmark{1}, Daniele Di Pompeo\IEEEauthorrefmark{2}, Vincenzo Stoico\IEEEauthorrefmark{3}, Michele Tucci\IEEEauthorrefmark{4}}
\IEEEauthorblockA{\textit{Department of Information Engineering, Computer Science and Mathematics} \\
\textit{University of L'Aquila}\\
\{\IEEEauthorrefmark{1}vittorio.cortellessa,
\IEEEauthorrefmark{2}daniele.dipompeo,
\IEEEauthorrefmark{4}michele.tucci\}@univaq.it,
\IEEEauthorrefmark{3}vincenzo.stoico@graduate.univaq.it
}}

\maketitle

\begin{abstract}
Software quality estimation is a challenging and time-consuming activity, and models are crucial to face the complexity of such activity on modern software applications. 

One main challenge is that the improvement of distinctive quality attributes may require contrasting refactoring actions on an application, as for trade-off between performance and reliability. In such cases, multi-objective optimization can provide the designer with a wider view on these trade-offs and, consequently, can lead to identify suitable actions that take into account independent or even competing objectives.

In this paper, we present an approach that exploits the \nsga multi-objective evolutionary algorithm to search optimal Pareto solution frontiers for software refactoring while considering as objectives: i) performance variation, ii) reliability, iii) amount of performance antipatterns, and iv) architectural distance. The algorithm combines randomly generated refactoring actions into solutions (i.e., sequences of actions) and compares them according to the objectives. 

We have applied our approach on a train ticket booking service case study, and we have focused the analysis on the impact of performance antipatterns on the quality of solutions. Indeed, we observe that the approach finds better solutions when antipatterns enter the multi-objective optimization. In particular, performance antipatterns objective leads to solutions improving the performance by up to 15\% with respect to the case where antipatterns are not considered, without affecting the solution quality on other objectives.
\end{abstract}

\begin{IEEEkeywords}
multi-objective optimization, software performance, software reliability, antipatterns, refactoring.
\end{IEEEkeywords}

\input{intro}
\input{background}
\input{approach}
\input{result}
\input{related}
\input{conclusion}

\balance
\bibliographystyle{IEEEtran}
\bibliography{biblio}

\end{document}

%% file: intro.tex
\section{Introduction}\label{sec:introduction}

In the last decades, multi-objective optimization techniques have been successfully applied to many model-driven software development problems~\cite{Ramirez:2018uz,Mariani:2017jd,Ouni:2015db,Ouni:2017db,Ray:2014ip,Bavota:2014kr,Kessentini:2012cb}. These techniques are evidently more effective on problems whose objectives can be expressed through quantifiable metrics. Problems related to non-functional aspects undoubtedly fit into this category, as witnessed by the vast literature in this domain~\cite{Aleti:2013gp,Martens:2010bn,DBLP:conf/mompes/AletiBGM09}. Most approaches are based on evolutionary algorithms~\cite{DBLP:journals/csur/BlumR03} that allow exploring the solution space by (re-)combining solutions. 

Software refactoring is a task that can be triggered by different causes, such as the introduction of additional requirements, the adaptation to new execution contexts, or the degradation of non-functional properties. The identification of optimal refactoring actions is a non-trivial task, mostly due to the large space of solutions, and there is still lack of automated support in this context.
A common aspect of multi-objective optimization approaches applied to model-driven software refactoring problems is that they search among design alternatives (\eg through architectural tactics~\cite{Koziolek:2011cg,NI2021106565}).

In this paper, we present an approach based on a multi-objective evolutionary algorithm (\ie \nsga) that searches sequences of refactoring actions to be applied on UML software models, leading to the optimization of several objectives, namely: i) performance variation (analyzed through Layered Queueing Networks), ii) reliability (analyzed through a closed-form model), iii) number of performance antipatterns (automatically detected) and iv) architectural distance. 
%
In particular, we are interested in studying the impact of performance antipatterns on the quality of refactoring solutions. 
Since it has been shown that removing performance antipatterns leads to systems that show better performance than the ones affected by them~\cite{DBLP:conf/cmg/SmithW01a,Smith:2003wv, DBLP:journals/infsof/ArcelliCP18}, we aim at studying if this phenomenon also holds in the context of muti-objective optimization, where performance improvement is not the only objective.



Since UML represents a commonly recognized standard in the software modeling domain, our approach applies to UML models augmented by MARTE \cite{MARTE} and DAM \cite{BernardiMP11} profiles, which allow to embed performance and reliability properties into UML models. However, UML does not provide native support for performance analysis, thus we introduce a novel model-to-model transformation that generates Layered Queueing Networks (LQN) from annotated UML models. The solution of LQN models feeds the performance variation objective.


Here we consider refactoring actions that are designed to improve performance in most cases. Since such actions may also have an impact on other non-functional properties, we introduce the reliability among the optimization objectives to study whether their combination could keep satisfactory levels of performance and reliability at the same time.
To quantify the reliability objective, we adopt an existing model for component-based software systems~\cite{CortellessaSC02} which can be generated from UML models.

In this context, a multi-objective optimization should also consider the distance between the initial model and the models resulting from applying refactoring actions.
Indeed, without an objective that minimizes such distance, the proposed solutions could be impractical because they would require to completely disassemble and re-assemble a model architecture. This is the rationale behind the inclusion of the architectural distance as an objective to minimize in our optimization. 

We recently investigated the effect of performance in a multi-objective optimization problem by presenting \easier~\cite{Arcelli:2018vo}, where we also observed the contribution of performance antipattern and architectural distance in the generation of Pareto frontiers. However, \easier relies on a performance-oriented architectural description language equipped with its own performance solver, thus limiting its application to more complex cases.

Summarizing, the main contributions of this paper are the following:

\begin{itemize} 
 \item we adopt UML as modeling notation, instead of DSLs that have been adopted in existing multi-objective approaches~\cite{Martens:2010bn,5949650,DBLP:conf/mompes/AletiBGM09,CORTELLESSA2021106568}, thus we move toward a widely adopted standard;
 
 \item we introduce a novel model-to-model transformation to generate Layered Queueing Networks from annotated UML models;
 
 \item we introduce an automated refactoring engine equipped with four refactoring actions that can be applied to UML models;
 
 \item we adopt a probabilistic concept of antipattern occurrence~\cite{DBLP:conf/fase/ArcelliCT15}, which allows us to analyze the sensitivity of solutions to the antipattern detection capability;
 
 \item we consider measures of performance, reliability, distance, and the number of performance antipatterns as competing objectives of our evolutionary process. 
 
\end{itemize}

We extensively applied our approach to the Train Ticket Booking Service~\cite{DBLP:conf/staf/Pompeo0CE19,DBLP:journals/tse/ZhouPXSJLD21} in order to answer the following three research questions:
\begin{itemize}
    \item \emph{RQ1}: Does antipattern detection contribute to find better solutions compared to the case where antipatterns are not considered at all?
    \item \emph{RQ2}: Does the probabilistic nature of fuzzy antipatterns detection help to include higher quality solutions in Pareto frontiers with respect to deterministic one?
    \item \emph{RQ3}: Is the approach able to keep the system reliability satisfactorily high?
\end{itemize}

\noindent Indeed, the experimentation lasted approximately 80 hours and generated more than 31,000 solutions.
Our results show that, by considering the reduction of performance antipatterns as an objective, we are able to obtain refactoring solutions that improve the performance by up to 15\% with respect to the case where antipatterns are not considered.

The structure of the paper is the following: \secref{sec:background} introduces basic concepts, \secref{sec:approach} describes the approach, in \secref{sec:results} we evaluate our approach on the case study and discuss the results and threats to validity. \secref{sec:related} reports related work, and \secref{sec:conclusion} concludes the paper.


%% file: background.tex
\section{Background}\label{sec:background}
We identify the four competing objectives of our evolutionary approach as follows: \perfq is a performance quality indicator that quantifies the performance improvement/detriment between an initial model and one obtained by applying the refactoring actions of a solution (\secref{sec:background:perfq}); \reliability is a measure of the reliability of the software system  (\secref{sec:background:reliability}); \pas is a metric that quantifies the amount of performance antipattern occurrences while considering the intrinsic uncertainty rising from thresholds used by the detection mechanism (\secref{sec:background:pas}); \achanges represents the distance between an initial model and one obtained by applying the refactoring actions of a solution (\secref{sec:background:distance}).

We employ the Non-dominated Sorting Algorithm II (\nsga) as our genetic algorithm~\cite{Deb:2002ut}, since it is extensively used in the software engineering community, \eg~\cite{Koziolek:2011cg,Mansoor:2015bm}. 
\nsga randomly creates an initial population from that the offspring population is created by applying the \emph{Crossover} operator with probability $P_{crossover}$, the \emph{Mutation} operator with probability $P_{Mutation}$. The union of the initial and the offspring populations is sorted by the \emph{Non-dominated sorting} operator, which identifies different Pareto frontiers with respect to considered objectives. Finally, the \emph{Crowding distance} operator cuts the worse half the sorted union off, and the remaining population becomes the initial population for the next step.

\subsection{Performance Quality Indicator (\perfq)}\label{sec:background:perfq}

\begin{table*}[htbp]
    \caption{Detectable performance antipatterns in our approach.}
    \label{tab:supported-pas}
    \centering
    \begin{tabular}{>{\raggedleft\arraybackslash}p{0.21\linewidth} >{\raggedright\arraybackslash}p{.75\linewidth}}
    \toprule
        Performance antipattern & Description \\
    \midrule
        Pipe and Filter              & Occurs when the slowest filter in a ``pipe and filter'' causes the system to have unacceptable throughput. \\
        Blob                         & Occurs when a single component either i) performs the greatest part of the work of a software system or ii) holds the greatest part of the data of the software system. Either manifestation results in excessive message traffic that may degrade performance. \\
        Concurrent Processing System & Occurs when processing cannot make use of available processors. \\
        Extensive Processing         & Occurs when extensive processing in general impedes overall response time.\\ 
        Empty Semi-Truck             & Occurs when an excessive number of requests is required to perform a task. It may be due to inefficient use of available bandwidth, an inefficient interface, or both. \\
        Tower of Babel               & Occurs when processes use different data formats and they spend too much time in convert them to an internal format. \\
    \bottomrule
    \end{tabular}
\end{table*}

\perfq quantifies the performance improvement/detriment between two models, and it is defined as follows:

\[\perfq(M)=\frac{1}{c}\sum\limits_{j=1}^{c} p_j\cdot \frac{F_j-I_j}{F_j+I_j}\]

\noindent where $M$ is a model obtained by applying a refactoring solution to the initial model, $F_j$ is the value of a performance index in $M$, and $I_j$ is the value of the same index on the initial model. $p\in\{-1,1\}$ is a multiplying factor that holds: i) $1$ if the $j$--th index has to be maximized (i.e., the higher the value, the better the performance), like the throughput; ii) $-1$ if the $j$--th index has to be minimized (i.e., the smaller the value, the better the performance), like the response time. 

Notice that, for performance measures representing utilization, $p$ also holds $1$ but we deﬁne a \emph{utilization correction factor} $\Delta_j$ to be added to each j–th term above, as defined in~\cite{Arcelli:2018vo}. The utilization correction factor penalizes refactoring actions that push the utilization too close to 1, i.e., its maximum value.

Finally, the global \perfq is computed as the average across the number $c$ of performance indices considered in the performance analysis.

As mentioned in the introduction, in order to obtain performance indices of a UML model, the analysis has to be conducted on a performance modeling notation, and we have adopted the Layered Queueing Networks (LQNs)\footnote{\url{http://www.sce.carleton.ca/rads/lqns/LQNSUserMan-jan13.pdf}} for this goal.
LQNs have been introduced to give a layered structure to Queuing Networks~\cite{franks_enhanced_2009}. Each layer contains software resources providing different classes of service. These elements are named \emph{tasks}, while classes of service are identified by \emph{entries}. Each task implements a queue that is served by a \emph{processor}. The behavior of tasks is modeled through requests among entries, or it is specified using \emph{activities}. An activity is the atomic unit of computation in an LQN. Activities can be related using ordering operators. An ordering operator indicates precedence between activities. \figref{fig:mapping} shows a graphical example of an LQN. 

An LQN is solved by decomposing it in several conventional Queuing Network submodels related to each other, one for each layer of the whole model. Each submodel is in turn made of two layers: the top layer includes tasks interpreting the clients, while tasks in the bottom layer are intended to be servers of the top ones.

\subsection{Reliability model}\label{sec:background:reliability}

The reliability model that we adopt here to quantify the \reliability objective is based on the model introduced in~\cite{CortellessaSC02}. The mean failure probability $\theta_S$ of a software system $S$ is defined by the following equation:

\[ \theta_S = 1 - \sum\limits_{j=1}^K p_j \left( \prod\limits_{i=1}^N (1 - \theta_i)^{InvNr_{ij}} \cdot \prod\limits_{l=1}^L (1 - \psi_{l})^{MsgSize(l,j)} \right) \]

This model takes into account failure probabilities of components ($\theta_i$) and communication links ($\psi_{l}$), as well as the probability of a scenario to be executed ($p_j$). Such probabilities are combined to obtain the overall reliability on demand of the system ($\theta_S$), which represents how often the system is not expected to fail when its scenarios are invoked.

The model is considered to be composed of $N$ components and $L$ communication links, whereas its behavior is made of $K$ scenarios. The probability ($p_j$) of a scenario $j$ to be executed is multiplied by an expression that describes the probability that no component or link fails during the execution of the scenario. This expression is composed of two terms: $\prod_{i=1}^N (1 - \theta_i)^{InvNr_{ij}}$, which is the probability of the involved components not to fail raised to the power of their number of invocations in the scenario (denoted by $InvNr_{ij}$), and $\prod_{l=1}^L (1 - \psi_{l})^{MsgSize(l,j)}$, which is the probability of the involved links not to fail raised to the power of the size of messages traversing them in the scenario (denoted by $MsgSize(l,j)$). 

\subsection{Performance Antipatterns}\label{sec:background:pas}

A performance antipattern describes bad design practices that might lead to performance degradation in a system. Smith and Williams have introduced the concepts of performance antipatterns in~\cite{DBLP:conf/cmg/SmithW01a}. These textual descriptions were later translated into 
first-order logic (FOL) equations~\cite{DBLP:journals/sosym/CortellessaMT14}.

A performance antipattern FOL is a combination of multiple literals, where each one represents a system aspect (\eg the number of connections among components). 
These literals must be compared to thresholds in order to reveal the occurrence of a performance antipattern. The identification of such thresholds is a non-trivial task, and using deterministic values may result in an excessively strict detection where the smallest change in the value of a literal determines the occurrence of the antipattern. For these reasons, we use the fuzzy threshold concept that has been introduced in~\cite{DBLP:conf/fase/ArcelliCT15}. An example of a performance antipattern fuzzy threshold is the following:

\[1 - \frac{UB(literal) - literal}{UB(literal) - LB(literal)}\]

The upper (UB) and the lower (LB) bounds, in the above equation, are the maximum and minimum values of the $literal$ computed on the entire system. 
Instead of detecting a performance antipattern in a deterministic way, such thresholds lead to assign probabilities to antipattern occurrences. 

In this study we detect the performance antipatterns listed in Table~\ref{tab:supported-pas}.

\subsection{Architectural distance}\label{sec:background:distance}

The architectural distance \achanges quantifies the distance of the model obtained by applying refactoring actions to the initial one. 
On one side, a \emph{baseline refactoring factor (BRF)} is associated with each refactoring action in our portfolio, and expresses the refactoring effort to be spent when applying the action.
On the other side, an \emph{architectural weight (AW)} is associated with each model element on the basis of the number of connections to other elements in the model.
Hence, the effort needed to perform a refactoring is quantified as the product between the \emph{baseline refactoring factor} of an action and the \emph{architectural weight} of the model element on which that action is applied.
\achanges is obtained by summing the efforts of all refactoring actions contained in a solution.

As an example, let us assume that a refactoring sequence is made up of two refactoring actions: A1 with $BRF(A1)=1.23$, and A2 with $BRF(A2)=2.3$. For each refactoring action the algorithm randomly selects a target element in the model. For instance, let those target elements be: E1 with $AW(E1)=1.43$, and E2 with $AW(E2)=1.32$. The resulting \achanges of A1 and A2 would be:

\[\achanges(A1,A2) = 1.23 \cdot 1.43 + 2.3 \cdot 1.32 \]

Details about the \emph{baseline refactoring factor} for each considered refactoring action are in \secref{sec:results:settings}.

%% file: approach.tex
\section{Approach}\label{sec:approach}

\begin{figure}[htb]
  \centering
  \includegraphics[width=0.8\linewidth]{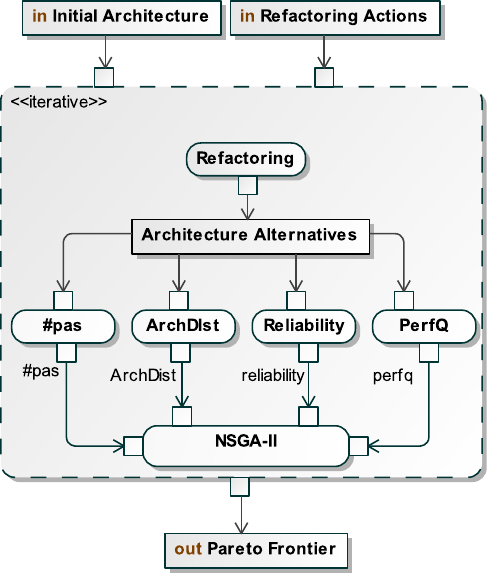}
  \caption{Our multi-objective evolutionary approach}
  \label{fig:process}
\end{figure}

Our process starts from an initial UML model and, through refactoring actions, generates alternative models that are functionally equivalent to the initial one, but that show different values for the non-functional attributes considered as objectives. Once an initial population is available, the evolutionary process starts seeking the optimal Pareto frontier. 
\figref{fig:process} is a graphical representation of the approach we present in this paper.


\subsection{Assumptions on UML models}

In our approach, we consider UML models including three views, namely \emph{static, dynamic} and \emph{deployment} views. The static view is modeled by a UML Component diagram in which static connections among components are represented by interface realizations and their usages. The dynamic view is described by UML Use Case and Sequence diagrams. A Use Case diagram defines user scenarios, while a Sequence diagram describes the behavior inside a single scenario through component operations (as defined in their interfaces) and interactions among them. A Deployment diagram is used to model platform information and map Components to Deployment Nodes. As mentioned before, we use an augmented UML notation by embedding two existing profiles, namely MARTE~\cite{MARTE} that expresses performance concepts, and DAM~\cite{BernardiMP11} that expresses reliability concepts.

\subsection{The Refactoring Engine}\label{sec:approach:refactoring}

The automated refactoring of software models is a key point when evolutionary algorithms are employed in order to optimize some model attributes. For the sake of full automation of our approach, we have implemented a refactoring engine that applies refactoring actions on UML models.




Each solution that our evolutionary algorithm produces is a sequence of refactoring actions that, once applied to an initial model, leads to a model alternative that shows different non-functional properties. 
Since our refactoring actions are combined during the evolutionary approach, we exploit the feasibility engine that verifies in advance whether a sequence of refactoring actions is feasible or not~\cite{DBLP:conf/kbse/ArcelliCP18}. 

The engine is based on the Cinn\'eide and Nixon~\cite{Cinneide:2000vh} framework. Each refactoring action is equipped with a pre- and post-condition, which describe the state of the subject model before and after the application of the action. The engine reduces a sequence of refactoring actions to a single refactoring action, which includes all the changes (see Equation~\eqref{eq:refFeasibility}).
For example, considering two refactoring actions ($M_{i}$, and $M_{j}$), then the global pre-condition is obtained by logical ANDing the first action pre-condition (${^{P_r}}M_{i}$) and all the parts of $M_{j}$ pre-condition that are not yet verified by $M_{i}$ post-conditions ($ M_{j}^{P_r}\ / \ M_i{^{P_o}} $) (see Equations~\eqref{eq:refPrecond}). Since the status of the model after a refactoring is synthesized by its post-condition, we can discard the parts of a subsequent refactoring pre-condition that, by construction, are already verified by its post-condition.
The global post-condition is obtained by logical ANDing all post-conditions within the sequence ($ M_{i}^{P_o} \wedge M_{j}^{P_o} $) (see Equation~\eqref{eq:refPost}).

\begin{subequations}
\begin{align}
    {^{P_r}}M_{i}^{P_o} \wedge {^{P_r}}M_{j}^{P_o} \longmapsto  {^{P_r}}M^{P_o} \label{eq:refFeasibility}\\
	{^{P_r}}M_{i} \wedge M_{j}^{P_r}\ / \ M_i{^{P_o}} \longmapsto {^{P_r}}M \label{eq:refPrecond}\\
	M_{i}^{P_o} \wedge M_{j}^{P_o} \longmapsto M^{P_o} \label{eq:refPost}
\end{align}
\end{subequations}


Our feasibility engine also allows to reduce the number of invalid refactoring sequences, thus reducing the computational time.
In the following, we report the four refactoring actions considered within the optimization process to generate model alternatives.

\paragraph{Clone a Node}
This action is aimed at introducing a replica of a Node. Adding a replica means that every deployed artifact and every connection of the original Node has to be in turn cloned. Stereotypes and their tagged values are cloned as well. The rationale of this action is to introduce a replica of a platform device with the aim of reducing its utilization.
%

\paragraph{Move an Operation to a new Component deployed on a new Node}
This action is in charge of randomly selecting an operation and moving it to a new Component. All the elements related to the moving operation (\eg links) will move as well. 
Since we adopt a multi-view model, and coherence among views has to be preserved, this action has to synchronize dynamic and deployment views. A lifeline for the newly created Component is added in the dynamic view, and messages related to the moved operation are forwarded to it. In the deployment view, instead, a new Node, a new artifact, and related links are created. The rationale of this action is to lighten the load of the original Component and Node. 
%

\paragraph{Move an Operation to a Component}
This action is in charge of randomly selecting and transferring an Operation to an arbitrary existing target Component. The action consequently modifies each UML Use Case in which the Operation is involved. Sequence Diagrams are also updated to include a new lifeline representing the Component owning the Operation, but also to re-assign the messages invoking the operation to the newly created lifeline. The rationale of this action is quite similar to the previous refactoring action, but without adding a new UML Node to the model. 
%

\paragraph{Deploy a Component on a new Node}
This action simply modifies the deployment view by redeploying a Component to a newly created Node.
In order to be consistent with the initial model, the new Node is connected with all other ones directly connected to the Node on which the target Component was originally deployed. The rationale of this action is to lighten the load of the original UML Node by transferring the load of the moving Component to a new UML Node. 
%


\subsection{UML to LQN Transformation}
As mentioned before, for a full automation of our approach we have implemented a model transformation (UML2LQN) that generates a Layered Queuing Network (LQN) starting from a UML model compliant with our assumptions. The performance metrics that are required from the evolutionary algorithm come out from the LQN analysis and are reported back in the UML model. The analysis is conducted using the LQN solver~\footnote{\url{https://github.com/layeredqueuing/V5}} provided by Franks \etal~\cite{lqn_solver_user_manual}. 

The transformation engine is based on the Extensible Platform for Integrated Languages for model management (Epsilon)~\cite{kolovos2010epsilon}. 

\begin{figure}
  \centering
  \includegraphics[width=\linewidth]{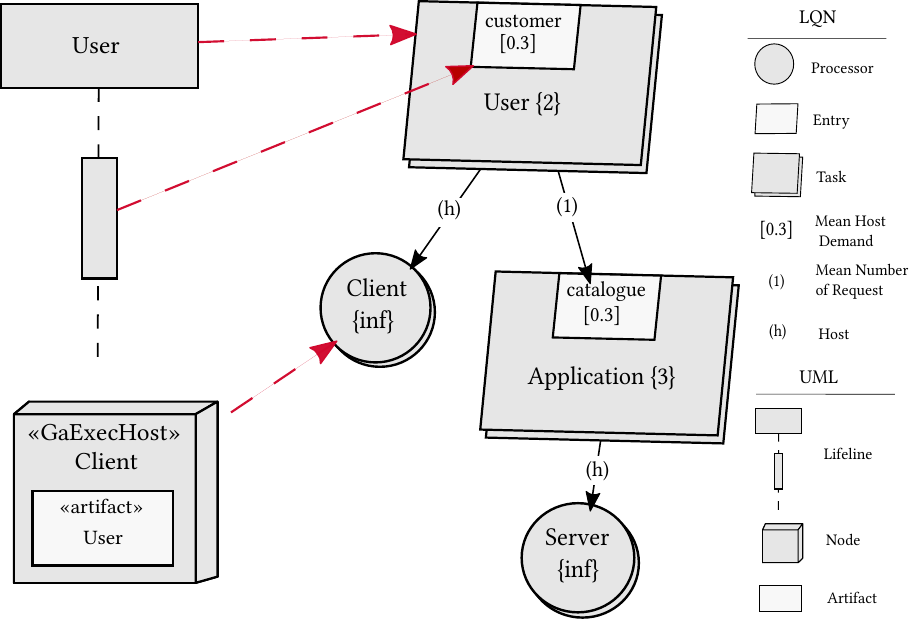}
  \caption{Graphical representation of the UML to LQN mapping.}
  \label{fig:mapping}
\end{figure}

UML2LQN mappings are listed in \tabref{tab:mapping}, and schematically exemplified in \figref{fig:mapping}. 
The first column of \tabref{tab:mapping} lists the UML Elements that are consumed by the transformation, while the third column lists the corresponding LQN Elements generated by the transformation. The second column, instead, lists the stereotypes that filter elements within the UML model. 

The transformation maps each UML Node to an LQN Processor. A node is included in the transformation if stereotyped with MARTE \emph{GaExecHost}. The transformation considers all nodes deploying at least an interacting component. A component is considered to be interacting if there is at least a UML Lifeline describing its behavior in at least one of the considered scenarios (\ie UML Use Cases).

Each Sequence Diagram should contain at least a lifeline representing the element (e.g., a user or another system) that triggers the scenario. This element is annotated as MARTE \emph{GaWorkloadEvent}. In our context, each scenario is activated by a UML Actor. Components and Actors are the elements that send and receive requests in the Sequence Diagram that describes the behavior originated by triggering a Use Case. For this reason, Components and Actors are mapped to LQN Tasks. Each lifeline has one or more units of behavior represented by a UML Behavior Execution Specification (BES). A BES represents a set of events. Each event stands for the reception or the emission of a message and is modeled by a UML Message Occurrence Specification (MOS). MOSs are associated with the requests made in an LQN, and consequently mapped to LQN \emph{synch-call}. It follows that BESs are mapped to LQN Entries.

A UML Message is transformed into an LQN Activity if stereotyped by MARTE \emph{GaStep}. Each LQN Entry holds at least an LQN Activity that models a synchronous call to a different LQN Entry. The \emph{GaStep.rep} tagged value denotes the number of requests of an LQN Activity. Instead, the \emph{GaStep.execTime} tagged value indicates the service time of a UML Message. This information is needed to calculate the demand for a call.

In an LQN, a request may be executed by multiple processors. UML2LQN exploits this feature by calculating the number of UML Nodes deploying the same component. These nodes are represented, in the obtained LQN, as a single processor having multiplicity larger than one.


Our approach is similar to the one from Altamini~\etal~\cite{altamimi_performance_2016}, but we could not adopt the latter because it starts from UML Activity Diagrams, whereas we introduce Sequence Diagrams to more explicitly express the interactions among components.  Both approaches report the results produced by the LQN solver back in the original UML model. In Altamini~\etal, this is accomplished using a trace model generated during the transformation. UML2LQN exploits an Epsilon Object Language (EOL) script that reads the solver output and directly introduces the results in the original UML model. The generation of a trace model would have added unnecessary complexity to the transformation. 
Hence, given our mapping, it is sufficient to use an EOL script that reports performance results back into the original model.

\begin{table}
\caption{The stereotype application on UML Elements and their mapping to LQN ones.}
\label{tab:mapping}
\begin{adjustbox}{width=\columnwidth}
\begin{tabular}{lll} 
\toprule
UML Element                      & Stereotype                    & LQN Element  \\ 
\midrule
Model                            & -                             & lqnmodel     \\ 
Node                             & MARTE::GaExecHost             & processor    \\ 
CommunicationLink                & DAM::DaConnector              & -            \\
Component                        & DAM::DaComponent              & task         \\ 
Actor                            & MARTE::GaWorkloadEvent        & task         \\ 
Behavior Execution Specification & -                             & entry        \\ 
Message                          & MARTE::GaStep                 & activity     \\ 
Message Occurrence Specification & -                             & synch-call   \\ 
Use Case                         & MARTE::GaScenario             & -            \\
\bottomrule

\end{tabular}
\end{adjustbox}
\end{table}

\subsection{Computing reliability on UML models}

The reliability parameters of the model introduced in \secref{sec:background:reliability} are annotated on UML models by means of the MARTE-DAM profile.
The probability of executing a scenario ($p_j$) is specified by annotating UML Use Cases with the \emph{GaScenario} stereotype. This stereotype has a tag named \emph{root} that is a reference to the first \emph{GaStep} in a sequence. We use the \emph{GaScenario.root} tag to point to the triggering UML Message of a Sequence Diagram and the \emph{GaStep.prob} to set the execution probability.
Failure probabilities of components ($\theta_i$) are defined by applying the \emph{DaComponent} stereotype on each UML Component and by setting, in the \emph{failure} tag, a \emph{DaFailure} element with the failure probability specified in the \emph{occurrenceProb} tag.
Analogously, failure probabilities of links ($\psi_{l}$) are defined in the \emph{failure.occurrenceProb} tag of the \emph{DaConnector} stereotype that we apply on UML CommunicationPath elements. Such elements represent the connection links between UML Nodes in a Deployment Diagram.
Sequence Diagrams are traversed to obtain the number of invocations of a component $i$ in a scenario $j$ (denoted by $InvNr_{ij}$ in our reliability model), but also to compute the total size of messages passing over a link $l$ in a scenario $j$ (denoted by $MsgSize(l,j)$). The size of a single UML Message is annotated using the \emph{GaStep.msgSize} tag. The Java implementation of the reliability model is available online\footnote{\url{https://github.com/SEALABQualityGroup/uml-reliability}}.

%% file: result.tex
\section{Evaluation}\label{sec:results}

In this section, we apply our approach to the Train Ticket Booking Service (TTBS) case study~\cite{DBLP:conf/staf/Pompeo0CE19,DBLP:journals/tse/ZhouPXSJLD21}. We report results of Pareto frontiers shapes, with a special focus on the role of performance antipatterns in the solution quality\footnote{Due to lack of space, we evaluated our approach on a single case study. However, its characteristics allowed a non-trivial validation exercise.}.

\subsection{Case Study}\label{sec:approach:running}

TTBS is a web-based booking application, and its architecture is based on the microservice paradigm. The system is made up of 40 microservices, and it provides different scenarios through users that can perform realistic operations, \eg book a ticket or watch trip information like intermediate stops. The application employs a docker container for each microservice, and connections among them are managed by a central pivot container.

Our UML model of TTBS is available online.\footnote{\url{https://github.com/SEALABQualityGroup/SEAA-replication-package/tree/main/case-study}}
The static view is made of \ttcomp, where each component represents a microservice. In the deployment view, instead, we consider \ttnode, each one representing a docker container.

Among all available TTBS scenarios shown in~\cite{DBLP:conf/staf/Pompeo0CE19}, in this paper we have considered \ttusecase, namely \emph{login}, \emph{update user details} and \emph{rebook}. We selected these three scenarios because they commonly represent performance-critical ones in a ticketing booking service. In particular,  each scenario is described by a UML Sequence Diagram. Furthermore, the model comprises two user categories: simple and admin users. The simple user category can perform the login and the rebook scenarios, while the admin category can perform the login and the update user details scenarios.

\subsection{Experimental setting}\label{sec:results:settings}

A configuration is defined by the combination of parameters related to the genetic algorithm, the performance antipattern detection, and the refactoring engine. In order to investigate which configuration produces better Pareto frontiers, we have executed multiple tuning runs to find a set of optimal configurations.

\begin{table}[htbp]
    \caption{The baseline refactoring factor for actions belonging to our portfolio.}
    \label{tab:action_brf}
    \centering
    \begin{tabular}{rl}
    \toprule
        Refactoring action & BRF \\
    \midrule
         Clone UML Node & 1.23\\
         Move Operation to a New Component to a new Node & 1.80\\
         Move Operation to a Component & 1.64\\
         Deploy a Component to a new Node & 1.45\\
     \bottomrule
    \end{tabular}
\end{table}

To set parameters related to the genetic algorithm, we have performed a tuning phase with the intent of increasing the quality of the Pareto frontiers. In particular, we have set the length of refactoring sequences to four actions, which represents a good compromise on the number of refactoring actions usually applied by a designer in a single session.
We have set the $P_{crossover}$ and $P_{mutation}$ probabilities to 0.8 and 0.2, respectively, following common configurations~\cite{DBLP:journals/ese/ArcuriF13}. The higher the values of these two probabilities, the greater the chance of generating an unfeasible sequence of refactoring actions, which in turn causes a longer simulation time due to a higher number of discarded sequences. For example, increasing $P_{crossover}$ could cause a lot of permutation among sequences, and it might lead to wrong or unfeasible sequences of refactoring actions.
The initial population size has been set to \textbf{16} elements (i.e., 16 different UML alternative models).
We have chosen this size for the initial population because it represents, for this case study, a good compromise between the diversity of solutions considered and the time required to create them.

Also, we have considered \{0.55, 0.80, 0.95\} as fuzziness thresholds to study the impact of performance antipatterns on Pareto frontiers. 
Regarding parameters related to refactoring actions, we have set the BRF of each refactoring action as reported in \tabref{tab:action_brf}. Since a tuning phase of the BRF is out of the scope of this paper, we did not investigate the impact of other BRF values on the overall quality. Thus, for the sake of the evaluation, we have chosen the values reported in the table, which represent the effort for applying those actions onto a model.


Furthermore, in order to reduce the randomness of the genetic algorithm, a variable number of multiple runs have been executed for each configuration. 
In the following, a Reference Pareto Frontier (RPF) refers to those solutions outcoming from the same problem configuration that are not dominated by other ones. 

Our experimental settings on the TTBS case study has generated \textbf{31,404} model alternatives and it has taken \textbf{82.4} hours of computation. We performed our experiments on a server equipped with two Intel Xeon E5-2650 v3 CPUs at 2.30GHz, 40 cores and 80GB of RAM. The replication package including the Java implementation of the approach, the case study and the results of our experimentation are available online\footnote{\url{https://github.com/SEALABQualityGroup/SEAA-replication-package}}.

\subsection{Results and discussion}


Results presented in this section are aimed at answering the aforementioned three research questions. 


\medskip

\RQ{RQ1}{Does antipattern detection contribute to find better solutions compared to the case where antipatterns are not considered at all?}

\noindent In order to answer this research question, we have conducted additional experimentation on the same problem configurations, where we have removed performance antipattern occurrences from the fitness function, thus reducing the optimization to the remaining three objectives. 

\begin{figure}[htb]
  \centering
  \includegraphics[width=\linewidth]{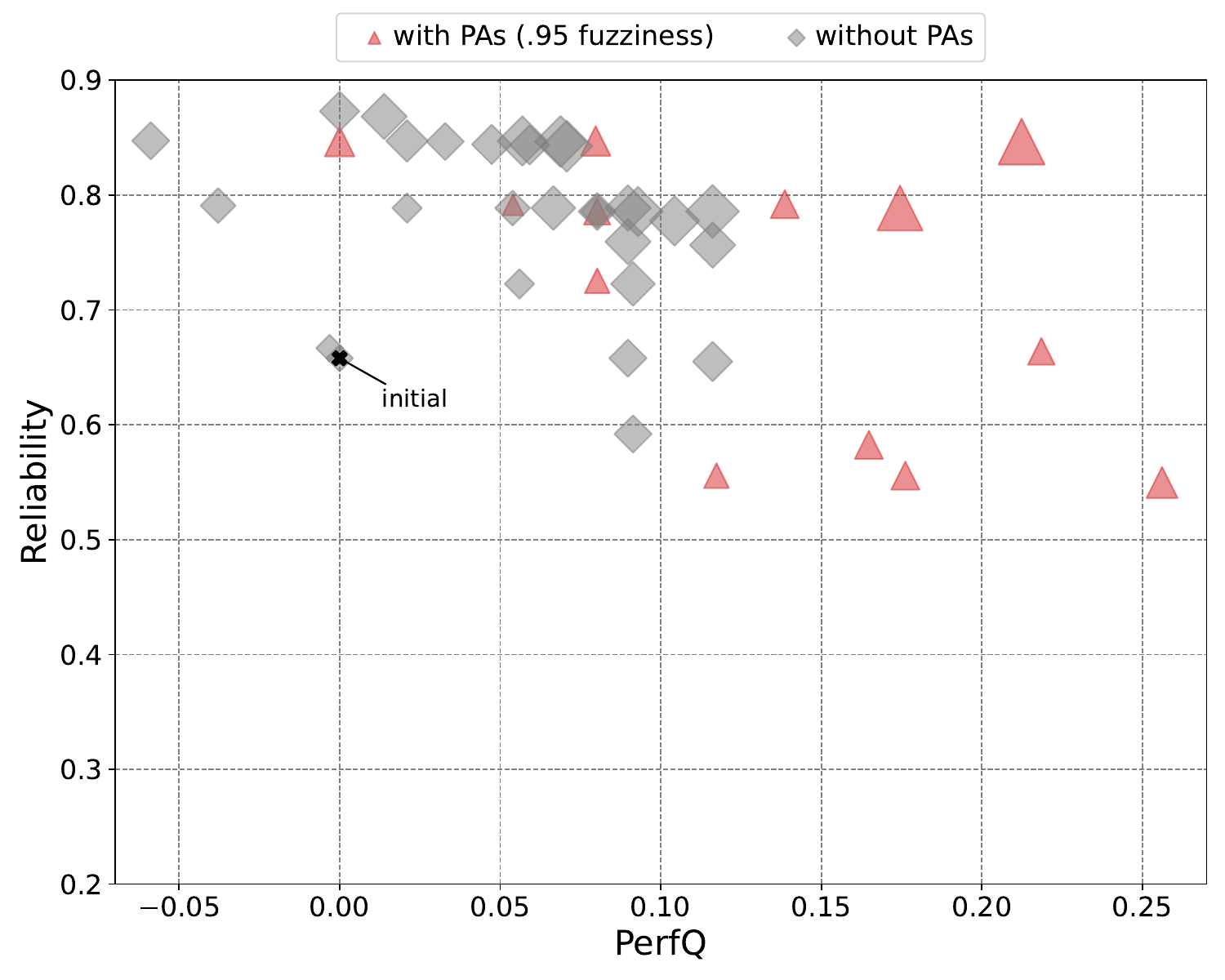}
  \caption{Reference Pareto Fronts obtained when considering performance antipatterns (with PAs) and without them (without PAs).}
  \label{fig:pas_vs_nopas_super}
\end{figure}

\figref{fig:pas_vs_nopas_super} reports the solutions in Reference Pareto Fronts across different problem configurations, where two objectives are represented on the axes and the architectural distance is represented by the symbol size of each solution. Hence, better solutions are the small sized ones located in the upper right corner of the figure. Two RPFs are reported in the figure: gray diamonds are solutions without PAs, whereas pink triangles with PAs, with a 0.95 fuzziness value. The initial model is identified by a black x. 

\figref{fig:pas_vs_nopas_super} provides clear evidence of higher quality solutions in the case with PAs, where antipattern detection helps to drive the evolutionary algorithm towards solutions that achieve \perfq improvements even twice higher than the ones without PAs (i.e., 0.25 vs 0.12). 


\answerRQ{On the basis of our experimentation, we can state that considering the performance antipattern occurrences in the optimization leads to better solutions than ignoring them.}



\bigskip

\RQ{RQ2}{Does the probabilistic nature of fuzzy antipatterns detection help to include higher quality solutions in Pareto frontiers with respect to deterministic one?}



\begin{figure}[htb]
\centering
    \centering
    \includegraphics[width=\linewidth]{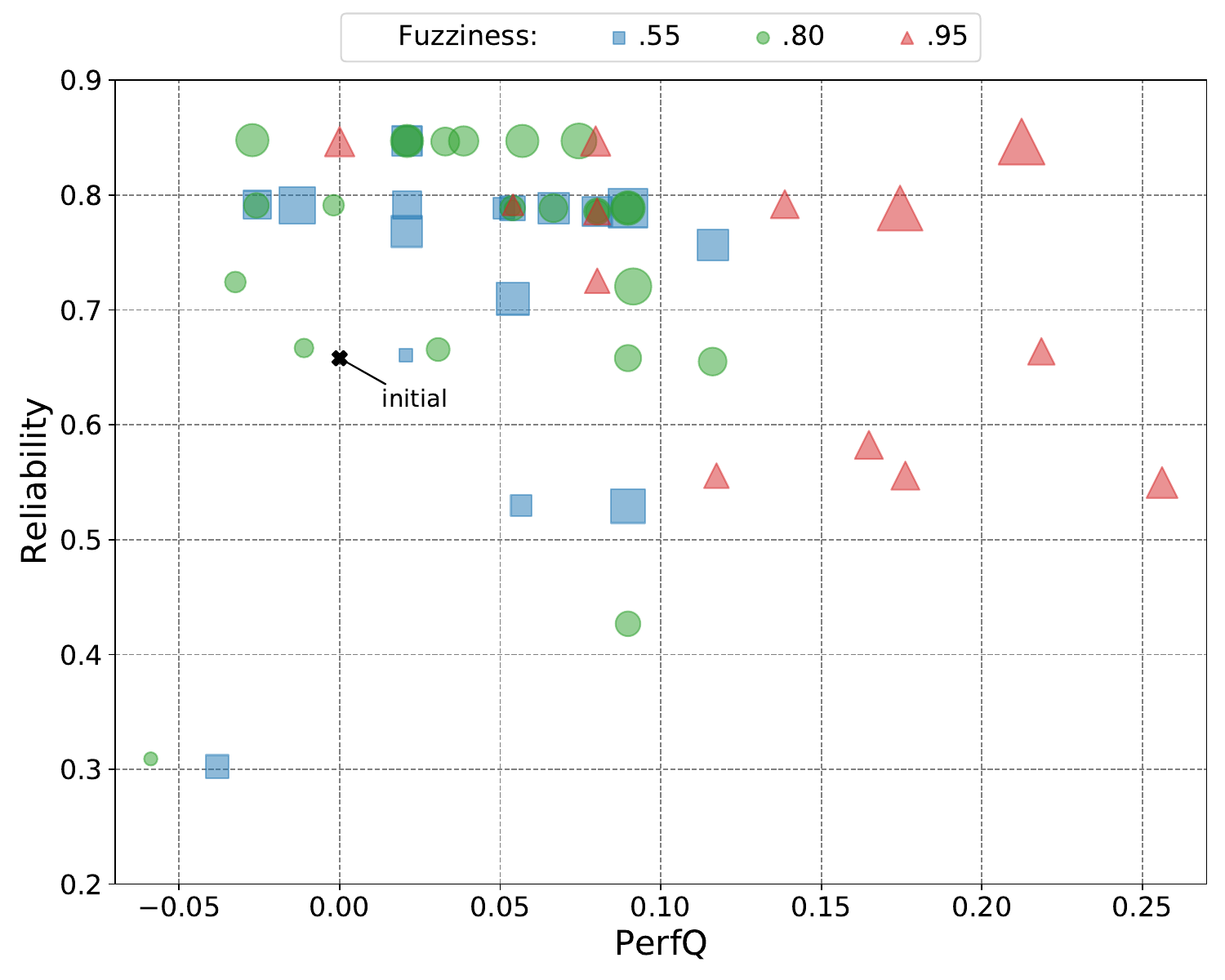}
  \caption{Reference Pareto Fronts obtained when varying the fuzziness threshold of the performance antipatterns detection.}
  \label{fig:fuzzy_th_102-super}
\end{figure}

\noindent In order to answer this research question, we run different problems while varying the values of the fuzziness threshold of the performance antipattern detection within \{0.50, 0.80, 0.95\}. The obtained RPFs are reported in \figref{fig:fuzzy_th_102-super}. 

The figure shows that moving from blue squared to green round solutions (i.e., in an increasing fuzziness value direction) already leads better solutions, at least in terms of reliability. By further moving from green round to pink triangle solutions it is possible to identify even better solutions, especially in terms of \perfq performance objective.

Hence, it seems that having a more deterministic antipattern detection (i.e., higher values of fuzziness) is better than a probabilistic one. However, the deterministic detection has the drawback of relying on fixed thresholds that must be computed in advance for each model alternative. For this reason, the trade-off between better quality solution and the effort to bind thresholds is likely domain-dependent and worth to be more investigated.

\answerRQ{On the basis of our experimentation, we can state that performance antipattern fuzzy detection does not help to improve the quality of Pareto frontiers.} 

\bigskip


\begin{table*}
\caption{Details of the values of \perfq, \reliability and \achanges obtained when varying the fuzziness threshold of the performance antipatterns detection.}
\label{tab:fuzz_vs_nopas_summary}
\centering
\begin{tabular}{|r|c|cccc|cccc|cccc|}
\cline{3-14}
\multicolumn{2}{c|}{} & \multicolumn{4}{c|}{\perfq} & \multicolumn{4}{c|}{Reliability} & \multicolumn{4}{c|}{\achanges} \\
\hline
Fuzziness & \# sol & min & max & median & mean  & min & max & median & mean       & min & max & median & mean \\
\hline
.55    & 16 &  -0.038 & 0.116 & 0.052 & 0.041 &  0.302335 &  0.847111 & 0.787196 &  0.713601 &  1.85 & 10.41 & 6.06 & 5.82 \\
.80    & 25 &  -0.058 & 0.116 & 0.053 & 0.040 &  0.309063 &  0.847808 & 0.788695 &  0.740425 &  1.97 &  9.18 & 4.83 & 4.96 \\
.95    & 13 &   0.000 & 0.256 & 0.138 & 0.134 &  0.549760 &  0.847111 & 0.785697 &  0.717795 &  3.03 & 14.67 & 5.49 & 6.56 \\
no PAs & 29 &  -0.058 & 0.116 & 0.066 & 0.056 &  0.591957 &  0.872976 & 0.788695 &  0.778037 &  2.46 &  9.75 & 5.40 & 5.88 \\
\hline
\end{tabular}
\end{table*}

\RQ{RQ3}{Is the approach able to keep the system reliability satisfactorily high?}

\figref{fig:fuzzy_th_102-super} shows that, regardless of the fuzziness value, most solutions improve reliability with respect to the initial model (i.e., the black x marker in the figure).
When looking at the mean values of reliability in Table~\ref{tab:fuzz_vs_nopas_summary}, it is clear that, even if in few cases the reliability of solutions decreases, on average we obtain an improvement from 8\% to 18\% (in absolute terms) with respect to the value of the initial solution, \ie 0.657925.
Moreover, a median higher than the mean also confirms that solutions with improved reliability are more likely to be part of a Pareto frontier.
In the best cases, the proposed refactoring actions are able to produce alternative models that achieve an improvement in reliability by over 28\% (in absolute terms).

\answerRQ{Our experiments show that, in the majority of cases, the approach is able to derive solutions with satisfactorily high reliability.}


\medskip
Table~\ref{tab:fuzz_vs_nopas_summary} reports a summary of the values of \perfq, \reliability and \achanges of solutions within different Pareto frontiers, both when performance antipatterns have been excluded and when a fuzzy detection is employed. In particular, in order to achieve the highest performance and reliability together, the algorithm has combined refactoring actions that, in some cases, have produced quite poor solutions in terms of architectural distance (\ie 14.67), as also shown by large triangles in the upper right corner of the figure. If, instead, we move towards lower values of reliability objective than the initial architecture one (\ie towards the lower right side of the figure), then the algorithm has generated closer models in terms of architectural distance in the range of \{1.85, 4.96\}, as also shown by small triangles in the $0.5$ to $0.7$ range on y-axis of \figref{fig:fuzzy_th_102-super}). Therefore, solutions that are able to achieve high performance and reliability usually show larger values of the distance metric, whereas the ones where either performance or reliability is pursued exhibit lower distance values.

\input{t2v}

%% file: t2v.tex
\subsection{Threats to validity}\label{sec:t2v}

We follow the Wholin et al.~\cite{wohlin2012experimentation} classification for the following threats.

\paragraph{Conclusion validity}
Our results might be affected by \emph{Conclusion validity} threats, since our considerations might change with better-tuned parameters for the \nsga. Also, problem parameter configurations might threat our conclusion. We did not perform an extensive tuning phase for the latter due to the long duration of each run, while we used common parameters for the \nsga, which should mitigate these threats.

\paragraph{Construct validity}
The way we have designed our problem and our experimentation might be affected by \emph{Construct validity} threats. In particular, the role played by the architectural distance objective on the combination of refactoring actions might affect the refactoring actions selection.

\paragraph{External validity}
Our approach might be affected by \emph{external validity} threats, because we have used a single modelling notation. We cannot generalize our results to other modelling notations, which could imply using a different portfolio of refactoring actions. In fact, the syntax and semantics of the modelling notation determine the amount and nature of refactoring actions that can be performed. We could mitigate these threats, for example, by using another modelling notation.

%% file: related.tex
\section{Related Work}\label{sec:related}

In the last decade, software architecture multi-objective optimization studies have been introduced to optimize various quality attributes (e.g., reliability, and energy \cite{Martens:2010bn,5949650,DBLP:conf/qosa/MeedeniyaBAG10,10.1007/978-3-642-13821-8_8}) with different degrees of freedom to modify the architecture (e.g., service selection~\cite{DBLP:conf/IEEEscc/RosenbergMLMBD10,Cardellini:2009:QRA:1595696.1595718}). A systematic literature review on architecture optimization can be found in~\cite{Aleti:2013gp}. We consider here as related work those approaches that directly involve multi-objective evolutionary algorithms and, on the other hand, approaches that exploit LQN as performance-oriented modelling notation \cite{DBLP:conf/wosp/WoodsidePPSIM05,DBLP:conf/qest/LiAZCP17,DBLP:conf/cascon/AltamimiP07,Koziolek:2011cg}.

\subsection{Software Architecture optimization}


Menasce \etal have presented a whole framework for architectural design and quality optimization~\cite{DBLP:conf/wosp/MenasceEGMS10}, where architectural patterns are used to support the searching process (e.g., load balancing, fault tolerance). Two limitations affects the approach: the architecture has to be designed in a tool-related notation and not in a standard modelling language (as we do in this paper), and it uses equation-based analytical models for performance indices that could be too simple to capture architectural details and resource contention.


Aleti \etal~\cite{DBLP:conf/mompes/AletiBGM09} have presented an approach for modeling and analyzing ADDL architectures~\cite{DBLP:books/daglib/0030032}. They have also introduced a tool aimed at optimizing different quality attributes while varying the architecture deployment and the component redundancy. Instead, our work relies on UML models and offers more complex refactoring actions as well as different target attributes for the fitness function. Besides, we investigate the role of performance antipatterns in the context of multi-objective software architecture refactoring optimization.

In the context of software architecture optimization, Cortellessa and Di Pompeo studied the sensitivity of multi-objective software architecture refactoring to configuration characteristics~\cite{CORTELLESSA2021106568}. They compared two genetic algorithms in terms of Pareto frontiers quality dealing with architectures defined in AEmilia, which is a performance-oriented ADL. In this paper, we change the modelling notation from AEmilia to UML, and we add the reliability as a new objective. Both approaches provide a refactoring engine, however, in this paper, the refactoring engine offers more complex refactoring actions since UML is more expressive than AEmilia.

\subsection{Layered Queueing Network approaches}


Koziolek \etal have presented PerOpteryx~\cite{Martens:2010bn,Koziolek:2011cg}, a performance-oriented multi-objective optimization problem. In PerOpteryx the optimization process is guided by tactics referring to component reallocation, faster hardware, and more hardware, which do not represent structured refactoring actions, as we intend to do in this paper. Moreover, PerOpteryx supports architectures specified in Palladio Component Model~\cite{Becker:2009cl} and produces, through model transformation, a LQN for of performance analysis. 


Rango \etal have presented SQuAT~\cite{10.1145/3132498.3132509}, an extensible platform aimed at including flexibility in the definition of an architecture optimization problem. SQuAT supports models conforming to Palladio Component Model language, exploits LQN for performance evaluation, and PerOpteryx tactics for architecture. 

\noindent We differ to PerOpteryx and SQuAt because we use UML as modelling notation. Therefore, our approach could be applied from a large audience of software designers since we rely on a standard notation instead of a specific language. 



Model-to-model (M2M) transformations from UML to LQN notations have been presented in~\cite{DBLP:conf/wosp/WoodsidePPSIM05,DBLP:conf/qest/LiAZCP17,DBLP:conf/cascon/AltamimiP07,altamimi_performance_2016}. In contrast with these approaches, we present a novel M2M transformation mapping that employs UML Sequence Diagrams as the behavioral view of software architectures instead of UML Activity Diagrams. Considering UML Sequence Diagrams has two benefits: they are adopted more frequently than UML Activity Diagrams during the software design~\cite{erickson2007can}, and they explicitly define method calls, while UML Activity Diagrams usually focus on workflows and processes. Therefore, our approach supports a more detailed behavioral representation in terms of time intervals between method calls.

%% file: conclusion.tex
\section{Conclusions and Future Work}\label{sec:conclusion}

In this work, we used \nsga to optimize UML models with respect to performance and reliability properties, as well as the number of detected performance antipatterns and the architectural distance.
We focused our study on the impact that performance antipatterns may have on the quality of refactoring solutions resulting from such optimization setting.

From our experimentation, we gathered interesting insights about the quality of the generated solutions and the role of perfomance antipatterns as an objective of the algorithm.
In this regard, we showed that, by including the detection of performance antipatterns in the optimization process, we are able to obtain better solutions in terms of performance and reliability.
Moreover, we also showed that, the more we increase the probability of detecting a performance antipattern using the fuzziness threshold, the better the quality of the refactoring solutions.
Another important aspect of our study was to ensure that the reliability of solutions remained within satisfactory levels.
In this respect, our experiments showed that we were in fact able to increase the reliability of refactoring solutions, with respect to the initial model, in the majority of cases.

As future work, we intend to tackle the threats to validity discussed before. In particular, we intend to investigate the influence of the experimental settings on the quality of Pareto frontiers.
Furthermore, we are interested in the role played by the architectural distance, and specifically in studying the effect of the baseline refactoring factor on the combination of refactoring actions.
In the same context, alternative models and additional parameters to evaluate the architectural distance can also be considered.
Similarly, for the reliability objective, we intend to include fault tolerance refactoring actions~\cite{DBLP:journals/infsof/CortellessaET20} in the refactoring portfolio, and to extend the reliability model to also take into account error propagation~\cite{DBLP:conf/cbse/CortellessaG07}.